\documentstyle[prl,
amsmath,amsfonts,aps,prc,epsf,amssymb,epsfig,multicol]{revtex}

\def\nn{\nonumber}

\begin{document}

\sloppy

\draft

\bibliographystyle{srt}

\title{Instant preheating mechanism and UHECR}
\author{ A. H. Campos$^{a,}$\footnote{hcampos@dfn.if.usp.br},
         J. M. F. Maia$^{b,}$\footnote{jmaia@cnpq.br}
         R. Rosenfeld$^{c,}$\footnote{rosenfel@ift.unesp.br}}

\address{\it  $^a$ Instituto de F\'{\i}sica - USP, Departamento de F\'{\i}sica Nuclear - CP 66318, S\~{a}o Paulo, SP,
Brazil} \vspace{1cm}

\address{\it  $^b$ CNPq - COCEX, SEPN 509, Bl. A, Ed. Nazir I, 70750-901, Bras\'{\i}lia, DF, Brazil} \vspace{1cm}

\address{\it  $^c$ Instituto de F\'{\i}sica Te\'orica - UNESP, Rua Pamplona, 145 - 01405-900, S\~{a}o Paulo, SP, Brazil}
\maketitle

\vspace{0.1cm}

\begin{abstract}
        Top-down models assume that the still unexplained Ultra High Energy
Cosmic Rays (UHECR's) are the decay products of superheavy
particles. Such particles may have been produced by one of the
post-inflationary reheating mechanisms and may account for a
fraction of the cold dark matter. In this paper, we assess the
phenomenological applicability of the simplest instant preheating
framework not to describe a reheating process, but as a mechanism
to generate relic supermassive particles as possible sources of
UHECR's. We use cosmic ray flux and cold dark matter observational
data to constrain the parameters of the model.
\end{abstract}

\vspace{0.2cm}

\noindent
PACS Categories:    98.70.Sa, 98.80.Cq

\vspace{0.2cm}


\begin{multicols}{2}
\relax

\section{Introduction}

        The possible observation of Ultra High Energy Cosmic Ray (UHECR's) events
with primary energies above $10^{20}$\,eV \cite{Agasa} constitute
one of the most intriguing puzzles in astroparticle physics (see,
for example, \cite{reviews}), although their origin and composition
are not yet understood. The usual bottom-up scenarios in which
particles should be accelerated by astrophysical objects do not
seem to provide a convincing solution to the puzzle. The arrival
direction of the primary particles should point to their sources
because at such energies the intergalactic magnetic field does not
deviate their direction of propagation. However, the clustering of
UHECR events observed in the available data is not statistically
significant and therefore there is no evidence that they arise from
point sources \cite{stat}. In addition, it would be necessary to
overestimate several parameters of such sources and their
acceleration regions in order to reach, marginally, the required
energies \cite{hillas}.

        The problem concerning cosmic ray sources is related to
the necessity that they must be located in our neighborhood, since
particles propagating at high energies suffer a rapid degradation
of their energy. For protons or nuclei as primaries, interactions
with the cosmic microwave background should cause a loss of their
energy due to photopion production. Such an effect should result in
a discontinuity in the cosmic ray spectrum for energies above $\sim
4\times10^{19}$ eV, the so-called Greisen-Zatsepin-Kuzmin~(GZK)
cutoff~\cite{GZK}. If the Auger Observatory data confirms that this
feature is not observed, it can be shown \cite{Aharonian} that
protons must have travelled less than $\sim 100$ Mpc (attenuation
length) in order to arrive at Earth with energies larger than
$10^{19}$ eV. The attenuation length for photons depends on their
initial energy and it is less than 100 Mpc for energies between
$10^{12}$ eV  to $10^{22}$ eV \cite{Protheroe}. Since neutrinos
have a very small cross section with nucleons within the Standard
Model, it seems difficult that they could produce air showers in
our atmosphere unless they had a yet unknown interaction
\cite{domokos}.

        In order to overcome such difficulties, another class of models
have been proposed \cite{berezinsky}. The primary particle would
not acquire kinetic energy continuously inside an accelerating
region (``bottom-up'' mechanism) as initially thought. Instead, the
highly energetic cosmic rays would be originated by the decay
products of superheavy particles of cosmological origin (``top
down'' mechanism). For simplicity, we will consider that such
particles have masses close to the GUT scale and would decay into
known particles, as quarks and leptons that evolve following the
QCD model \cite{topdown}. The quarks hadronize producing a small
fraction of nucleons and pions that in their turn decay into
photons, neutrinos and electrons and their corresponding
antiparticles. Therefore, from the decay of such a supermassive
particle it is possible to produce energetic photons, neutrinos and
leptons, together with a small percentage of nucleons. Depending on
which kind of particle is the primary, different attenuation
lenghts can be obtained, so that one can establish at what minimum
distances the supermassive particles sources should be located.

        There are different exotic candidates to play the main role in top-down
scenarios, such as decaying topological defects \cite{vilenkin} or
evaporating primordial black holes \cite{barrau}. The simplest top
down models (at least from the particle physics point of view)
involve supermassive metastable particles sometimes called
WIMPzilla's \cite{wimpzillas}. Due to their colossal masses, such
particles were presumably produced during the post-inflationary
epoch and could contribute to a part or to the whole of the dark
matter that accounts for about 30\% of the energy density of the
Universe. In order to explain the theoretically estimated UHECR's
fluxes \cite{Kuzmin}, such particles must be decaying now and have
to be located in our neighborhood, which is expected, assuming they
are concentrated in our galactic halo.

        In this work we study the possibility of producing
WIMPzillas in the post-inflationary process called instant
preheating, suggested by Felder, Kofman and Linde (FKL), originally
proposed as an alternative preheating mechanism~\cite{felder}. 
This process seems to be essential for particle production in models of 
quintessential inflation \cite{HAR,SS}.

In such a scenario, scalar particles $\chi$ are non-perturbatively
produced from the coherent oscillations of the inflaton $\phi$,
have their masses ``boosted'' due to their coupling to the field
$\phi$, and subsequently decay into supermassive metastable
fermions $\psi$. The idea of examining stable supermassive
particles in this context was addressed by Felder {\it et al.}, but its
consequences either as dark matter or as cosmic rays primaries were
not calculated in detail.

        More specifically, there is a relation between the density
parameter of these particles and their lifetime. If
$\psi$-particles compose the whole of the dark matter
($\Omega_{\psi}\equiv {m_{\psi} n_{\psi} \over \rho_{crit}} \sim
0.3$) \cite{Freedman}, a maximum lifetime limit will be found. On
the other hand, a lower limit on the abundance of such
particles can be obtained if lifetimes are constrained to be larger
than the age of the Universe ($\tau_{\psi}
\geq10^{10}$ yr) \cite{leticia}. As will be shown later,
such limits impose severe constraints on the parameters of the FKL
mechanism.

        This paper is organized as follows. In the next section we review
some features of the nonperturbative processes more directly
related to the production of massive scalar particles. In Section
III we perform a detailed calculation of superheavy $\psi$ particle
production, extending the previous results in \cite{felder}. In
Section IV we discuss our main results for produced particles
considered as dark matter in our current Universe and present the
parameter space for this model, which is in accordance with
cosmological data. In the last section we present a summary of our
main results and discuss their consequences.

\section{Production of $\chi$-particles}

        After inflation, matter had to be (re)created, since the Universe
became empty. In the case of chaotic inflation, the scenario
considered here \cite{chaotic}, particle production may occur
during the quasiperiodic evolution phase of the inflaton field. To
study such a period we assume the lagrangian:
\begin{equation}
L= \frac{1}{2} \partial_\mu \phi \partial^\mu \phi - V(\phi) +
\frac{1}{2} \partial_\mu \chi \partial^\mu \chi -
\frac{1}{2} m_\chi^{ 2}\chi^2
-\frac{1}{2} g^2\phi^2\chi^2.
\label{lagrangian}
\end{equation}
The inflaton field $\phi$ produces quantum scalars bosons $\chi$ with bare
masses $m_\chi$ through the interaction term $-g^2 \phi^2
\chi^2/2$. For simplicity, we will limit our analises to the
quadratic potential $V(\phi)=m_{\phi}^2
\phi^2/2$ (the simplest chaotic inflation model) with $m_\phi
\approx 10^{-6} M_{Pl}$, as required by microwave background
anisotropy measurements.

       During inflation we can neglect the contribution of the $\chi$ field and
the equation of motion for the $\phi$ field is given by:
\begin{equation}
\ddot{\phi}+3H\dot{\phi}+m_{\phi}^2 \phi=0,
\label{eqmotphi}
\end{equation}
where $H=\dot{a}/a$ is the Hubble parameter and obeys the Friedmann equation:
\begin{equation}
H^2=\frac{8\pi}{3M_{Pl}^2} \left ( \frac{\dot{\phi}^2}{2} +
\frac{m_{\phi}^2 \phi^2}{2}\right).
\label{friedmann}
\end{equation}

        As far as the slow roll conditions are valid ($\ddot{\phi} \ll
3H\dot{\phi}$, $\dot{\phi}^2/2 \ll V(\phi)$), $\phi \approx M_{Pl}/3$), the
Universe is in an inflationary phase. Right after inflation, the $\phi$
field oscillates about the minimum of its potential, with decreasing
amplitude due to the damping term $3H\dot{\phi}$, and the solution of the
above equation becomes:
\begin{equation}
\phi(t) \approx \frac{M_{Pl}}{3} \frac{\sin{\left(m_\phi t\right)}}{m_\phi t}
\label{sol.phi-pre}.
\end{equation}

       The $\phi$ field may produce $\chi$-particles
during this oscillating phase due to nonperturbative effects
\cite{brandenberger,kofminho}, provided the coupling constants have
appropriate values. As $\chi$ particles are bosons, such a process
may lead to an explosive particle production through parametric
resonance of the $\chi$ field \cite{kofmao}. To illustrate
this behavior, we consider the quantum nature of $\chi$
\begin{equation}
\stackrel{\wedge}{\chi}(t,\vec{x})= \frac{1}{(2\pi)^{3/2}}
\int{d^3k(\stackrel{\wedge}{a}_k\chi_k(t)e^{-i\vec{k}\cdot\vec{x}} +
          \stackrel{\wedge}{a}^\dag_k\chi_k^\ast(t)e^{i\vec{k}\cdot\vec{x}})},
\label{operador-chi}
\end{equation}
where $\stackrel{\wedge}{a}_k$ e $\stackrel{\wedge}{a}^\dag_k$ are
the creation and annihilation operators, respectively. The
equations of motion for the $k-$modes of the $\chi$ eigenfunctions
are given by:
\begin{equation}
\ddot{\chi}_k(t) + 3H\dot{\chi}_k(t) + \left( \frac{k^2}{a^2(t)} +
 m_\chi^ 2 + g^2\phi^2(t) \right) \chi_k(t) =0.
\label{eq.movimento-chi}
\end{equation}
Rewriting the above equation in terms of a more convenient variable
$X_k \equiv a^{3/2} \chi_k$, we obtain:
\begin{equation}
\ddot{X_k} + \left(\frac{k^2}{a^2} + m_\chi^2 + g^2\phi^2 \right) X_k =0 ,
\label{eq.movimento-X}
\end{equation}
where we used the fact that, for the quadratic chaotic potential,
the inflaton coherent oscillations redshift as nonrelativistic
matter. Note that this is an oscillator equation with a variable
frequency
\begin{equation}
 \omega_k(t)=\sqrt{\frac{k^2}{a^2(t)} + m_\chi^2 + g^2\phi^2(t)}.
\label{omega}
\end{equation}
The effective mass of $\chi$ is defined as
\begin{equation}
m_\chi ^{\rm eff}(t)=\sqrt{m_\chi^2 + g^2\phi^2(t)}.
\label{efetiva}
\end{equation}
Depending on the values of the parameters, the time variation of
$\omega_k$ will not be adiabatic. The loss of adiabaticity takes
place when $\phi$ field passes through the minimum of its
potential, the region where $\omega_k(t)$ changes more rapidly. In
such a case, there will be an inequivalence between the $X$ vacua
defined before and after the passage of the inflaton through the
minimum, which can be interpreted as production of $\chi$ particles
\cite{kofmao}. This particle production process has been considered
mainly for preheating proposals, since it happens before the usual
perturbative reheating. Alternatively, such a coupling between
$\phi$ and $\chi$ can be used in models with production of heavy
metastable particles in the early Universe. In such a case, the
produced particles may have masses larger than the inflaton mass
\cite{Chung}. In what follows, we work on the latter approach and
consider $g$ and $m_\chi$ as free parameters to be estimated from
the available cosmological data.

\section{The Felder-Kofman-Linde Mechanism}

        Felder {\em  et al.}$\cite{felder}$ noticed that the preheating
could be efficient even in a single $\phi$ field oscillation if the
$\chi$ particles produced at the minimum of $V(\phi )$ were allowed
to decay perturbatively into $\psi$ fermions when $\phi$ reached
the maximum of its potential energy. The decay of such fermions
into other particles and their subsequent thermalization could
complete the reheating process without the necessity of parametric
resonance. The authors also suggested the possibility of using such
a mechanism to produce heavy particles in order to explain cosmic
rays with energies above the GZK cutoff in a top-down approach.
Here, we perform explicit calculations following such a suggestion
by assuming that $\psi$ particles are produced
non-relativisticaly. In this way the energy transfer process from
the inflaton field is more efficient, since only a very ``fat"
$\chi$ particle can decay into $\psi$ particles.  The whole process
can be schematically represented as
\[
\phi \stackrel{g}{\longrightarrow} \chi\stackrel{g'}{\longrightarrow}\psi,
\]
where $g'$ is the coupling constant of a Yukawa interaction
$g' \overline{\psi}\psi\chi$
added to the lagrangian Eq.(\ref{lagrangian}) to account for the
interaction between $\psi$ and $\chi$.

        From the Yukawa interaction term, the decay rate
of $\chi$ bosons into $\psi$ fermions is
\begin{equation}
\Gamma(\chi \rightarrow \overline{\psi}\psi)={g^{\prime 2}m_\chi^{\rm eff}
\over 16\pi}\left[1-\left({2m_\psi \over m_\chi^{\rm eff}}\right)^2 \right]^{3 \over
2}.
\label{gama}
\end{equation}
Note that the above rate is not constant, since $m_\chi^{\rm eff}$
defined by (\ref{efetiva}) is time-dependent, and $\chi$ particles
tend to decay at large values of $\phi(t)$. This is the most
interesting feature of the model: a great amount of energy can be
transfered from $\phi$ to the stable fermions because the bosons
$\chi$ decay when their variable masses are at a maximum value.

        It is possible to obtain a relation between the couplings
$g$ and $g'$ if we take the interval $\Delta t=t_f-t_i$ around the
maximum value of $\phi$ evolution for which $m_\chi^{\rm eff}$ is
large enough to allow the decay of $\chi$ particles into
nonrelativistic $\psi$ fermions. Denoting by $n_i$ and $n_f$  the
number densities of $\chi$ before and after the passage of $\phi$
through the local minimum of $V(\phi)$, repectively, we have that
\begin{equation}
\ln\left({n_{f} \over n_{i}}\right)=-\int_{t_i}^{t_f} \Gamma(t) dt.
\label{taxa-t}
\end{equation}
It is convenient to work with a more intuitive time variable, i.e.
the number of oscillations
\begin{equation}
N={m_\phi t \over 2\pi}
\label{N(phi)}
\end{equation}
of $\phi$. In such a case, the solution (\ref{sol.phi-pre}) for
$\phi$ becomes
\begin{equation}
\phi(N) \approx \frac{M_{Pl}}{3} \frac{\sin{(2\pi N)}}{2 \pi N}.
\label{sol.phi-preII}
\end{equation}
This is a good parametrization from $N=0.25$ and later if we fix
$N=0.5$ to be the instant when the inflaton field crosses the
minimum of its potential for the first time. We will consider that
the maximum momentum of $\psi$ particle to be its mass, that is,
$p_{\psi} \lesssim m_{\psi}$. The maximum available energy for the
creation of a $\psi$ pair is $g\Phi$, where $\Phi$ is the amplitude
of $\phi(N)$.  This determines the largest
interval $\Delta N$ around the maximum amplitude of the $\phi$
field, $N \approx 0.72$ ($\Phi \approx 0.07 M_{Pl}$), for which $m_\chi^{\rm eff} \approx g
\Phi$ is large enough to allow a pair creation. Such an interval is
found to be $\Delta N=0.25$. Additionally, we require that
approximately 90\% of the $\chi$ particles decay during this
interval. From Eq. (\ref{taxa-t}) we obtain,
\begin{equation}
\ln\left({10 \over 100}\right) \lesssim - \frac{1}{2^{3/2}} \times {g^{\prime 2} g \over 16\pi}\int_{0.60}^{0.85}
 {2\pi \over m_\phi} \phi(N) dN,
\end{equation}
with $\phi(N)$ given by (\ref{sol.phi-preII}). Solving numerically
the above integral, we obtain that
\begin{equation}
g^{\prime 2}g \gtrsim 3 \times 10^{-3}.
\label{limite}
\end{equation}
This upper limit is almost one order of magnitude larger than the
Felder {\it et al} estimate \cite{felder} ($g^{\prime 2} g \approx
5 \times 10^{-4}$), since they did not consider the phase space
factor in the expression for the decay rate (\ref{gama}). It is
important to notice that the only arbitrariness in our assumptions
is the fraction of remaining $\chi$ particles (90\%), but for 
reasonable choices (say $n_f/n_i$ between 1/2 and 1/100) the
constraints on $g$ and $g'$ do not vary appreciably. All the other
constraints are consequences of the assumption of a maximum
momentum $p_{\psi} \lesssim m_{\psi}$ so that $\psi$ may be considered
as non-relativistic. Naturally, the limits on $g^{\prime 2}g$ would
be even tighter if we had considered values for larger momenta.
Therefore, we are being conservative in our estimates. We also
verified that although one should assume the time dependence of the
decay rate on the above calculations, it does not bring any
important difference if compared to the estimate found in
\cite{felder} where a constant decay rate, $\Gamma$, was used. This seems
reasonable, since the integration interval for $N$ is taken to be
around the maximum of $\phi(N)$ where the sine function does not
vary significantly. Such care would be necessary if we were
studying the production of relativistic particles.

        Independently of the calculation details, our main goal in
this work is to find the largest possible region of the parameter
space of the FKL mechanism that is phenomenologically allowed by
the available data. Assuming that the $\psi$ fermions are the
metastable massive particles that we are looking for, we need to
evaluate their present abundance supposing that they contribute to
the energy density of the dark matter today. Since for each
decaying $\chi$ a $\overline{\psi}\psi$ pair is created, we have to
find the number density $n_\chi$ of the $\chi$ bosons for each
oscillation of the inflaton. For the first oscillation, $n_\chi$
can be calculated from the solution of Eq. (\ref{eq.movimento-X})
about the minimum of $V(\phi)$, so that the total number of
$\chi$-particles \cite{felder} is
\begin{equation}
n_\chi = {1 \over (2\pi)^3} \int d^3k \: n_k
       = {(g|\dot{\phi}_0|)^{3 \over 2} \over 8\pi^3} \exp \left(-{\pi m_{\chi}^2 \over
       g|\dot{\phi}_0|}\right),
\label{tot.part.number.ip}
\end{equation}
and $\dot{\phi}_0=m_\phi \Phi$ is the field velocity near the
minimum of the potential. The model describes the above production
of $\chi$ particles, the boosting of their masses and their
subsequent decay into $\psi$ WIMPzillas with masses
\begin{equation}
m_{\psi}\approx g\Phi \approx 0.07gM_{Pl}.
\label{eq:mpsi}
\end{equation}

        In order to verify that most of the $\psi$ particles will be produced in
the first oscillation of $\phi$, we compare the number density of
produced $\chi$ particles in the second passage by the minimum of
the $\phi$ potential, $n_\chi^{(2)}$, to the first one,
$n_\chi^{(1)}$. By taking into account the dilution of
$n_\chi^{(1)}$ due to the Universe expansion between the
consecutive passages, it is found that (for details see the
Appendix)
\begin{equation}
{n_\chi^{(2)}\over n_\chi^{(1)}}= \sqrt{2} \exp \left({-6\pi^2 \times 10^{6} \over g}
{m_\chi^2 \over M_{Pl}^2} \right).
\label{n2-n1}
\end{equation}
Therefore, the exponential term could provide the desired
suppression between the two first passages. In fact, it can be shown that
the ratio between two consecutive passages tends to a constant prefactor
multiplying the exponential, and, as long as the suppression is assured,
particle production will be negligible for all subsequent oscillations. We
will see in the next section that such suppression is verified, since for
typical values of the parameters, say,
$g=10^{-2}$ and $m_{\chi}= 10^{-4}M_{Pl}$, $n_{\chi}^{(2)}\approx 10^{-26}
n_{\chi}^{(1)}$ and hence it is reasonable to consider only the first 
passage in our calculations.

\section{$\psi$ abundance and Ultra High Energy Cosmic Rays}

        If the $\psi$ particles are the superheavy relics that decay into
the observed UHECR's, we can use the presumed cosmic ray flux
associated to their decays \cite{reviews} in order to estimate
limits on the parameters of the model discussed above.

        By accounting for the dilution of $\psi$ particles since their
production until today, one may find the associate density
parameter $\Omega_{\psi}(t_0)=\rho_{\psi}(t_0)/\rho_c$. In order to
obtain an expression for this parameter, we have to consider three
different moments in the history of the Universe: the production of
$\psi$ particles ($t_p$), the end of the reheating period
($t_{rh}$) and today ($t_0$). We can then write:
\begin{equation}
\frac{\rho_\psi(t_0) T_0}{\rho_R(t_0)} = \frac{\rho_\psi(t_{rh})
T_{rh}}{\rho_R(t_{rh})},
\label{ratio-rho}
\end{equation}
where we assumed thermal equilibrum for the relativistic components
and the fact that $\rho_\psi(t_0) a{_0}^3 =
\rho_\psi(t_{rh}) a{_{rh}}^3$. On the other hand we can suppose
that the Universe will be reheated from an instant convertion of
the remanescent inflaton  density energy into relativistic
particles, so that
\begin{equation}
\frac{\rho_{\psi}(t_{rh})}{\rho_{R}(t_{rh})} = \frac{\rho_{\psi}(t_{rh})}{\rho_{\phi}(t_{rh})}
= \frac{\rho_{\psi}(t_{p})}{\rho_{\phi}(t_{p})},
\end{equation}
where the last equality is obtained by considering that the $\psi$
and $\phi$ field coherent oscillations redshift as non relativistic
matter. Note that we are supposing that only the $\psi$ particle
production in the first passage is important (see expression
(\ref{n2-n1})).

Finally we can substitute $\rho_\phi(t_p)=(3M_{Pl}^2/8\pi) H_{p}^2$
into (\ref{ratio-rho}) so that
\begin{equation}
\Omega_\psi(T_0)h^2=\Omega_R(T_0)h^2\left({T_{rh} \over
T_0 }\right) {8\pi \over 3} {m_\psi n_\psi \over M_{Pl}^2H_p^2},
\label{abundance1}
\end{equation}
where $T_0\approx 2.7K$ is the present CMB temperature, $h \equiv
H/(100  \mbox{km}.\mbox{s$^{-1}$}.\mbox{Mpc$^{-1}$)}$ and
$\Omega_R(T_0)h^2\approx 4.3 \times 10^{-5}$ is the current
radiation density parameter. The other parameters are $H_p^2 =
(8\pi/3M_{Pl}^2)(\rho_{\phi_p})= (8\pi/3M_{Pl}^2)(m_\phi^2\Phi^2/2)
\approx (4\pi/3) \times 10^{-14} M_{Pl}^2$ \quad 
and, given that $n_{\chi} = n_{\psi}/2$ and for
nonrelativistic fermions $m_{\chi}^{\rm eff} \approx 2 m_{\psi}$,
we find that $m_{\psi}n_{\psi} \approx m_{\chi}^{\rm eff} n_{\chi}$, 
where $n_\chi$ is given by equation (\ref{tot.part.number.ip}). Choosing
$T_{rh}=10^9$ GeV, the above equation can be rewritten as
 \begin{eqnarray}
\Omega_\psi(T_0)h^2 & \approx  &  4.72 \times 10^{18} \left(T_{rh} \over {10^9 {\rm GeV}}\right)
     \left({2.7 {\rm K} \over T_0}\right) g^{3/2} \times \nn \\
      & &\sqrt{{m_\chi^2 \over M_{Pl}^2}+ 10^{-2}g^2}
     \: \exp {\left(-{\pi \times 10^7 \over g} {m_\chi^2 \over
     M_{Pl}^2}\right)}.
\label{omega-psi}
\end{eqnarray}

        If $\psi$ WIMPzillas are required to explain UHECR's, it is possible
to obtain a relation between their abundance ($\Omega_\psi$), their
masses ($m_\psi$), the lifetime of $\psi$ ($\tau_{\psi}$) and the
UHECR's fluxes (for details see references \cite{reviews,leticia}).
As we mentioned earlier, in order to obtain such relations it is
necessary to adopt a model in which particles are produced from the
$\psi$ decay. This topic is an issue by itself, and we will follow
the usual assumption of extrapolating QCD mechanisms valid in the
the quark-hadron fragmentation process for the higher energies
considered here. As we will show later on, the use of a specific
value for the flux will not alter the region of the allowed
parameter space appreciably. Such a choice implies that photons
dominate the primary spectrum by a factor of $\approx 6$ over
protons \cite{reviews}. This means that, if we consider the
observable ultra high energy cosmic ray fluxes as due to extremely
energetic photons resulting from the decay of $\psi$ particles, we
can write:
\begin{equation}
\tau_\psi=3.16 \times 10^{18}\;f\;(\Omega_\psi
h^2)\;\left(\frac{M_{Pl}}{m_\psi}\right)^{1/2},
\end{equation}
where $f$ measures the clustering of $\psi$ particles inside the
galactic halo. It is taken as 1 for a uniform distribution of
superheavy WIMP's, but can be considered as approximately  $10^3$
if they are concentrated in galactic halos (see \cite{reviews}). We
will assume the latter case in what follows, i.e. $f \approx
10^3$, so that following limits on the parameters of the model can be
established:
\vspace{0.5cm}

$\bullet$ if they constitute the whole of the dark matter (for
$\Omega_\psi \sim 0.3$, $h\sim 0.7$) \cite{Freedman})
\begin{equation}
\Omega_\psi(T_0)h^2 \sim 0.14;
\label{eq:upper}
\end{equation}

$\bullet$ if they have a lifetime of the order of the age of the
Universe
\begin{equation}
 \tau_\psi \approx 10^{10} \mbox{years} \rightarrow \Omega_\psi(T_0)h^2 \approx
 3.16 \times 10^{-12} \left(\frac{g\Phi}{M_{Pl}}\right)^{1/2}.
\label{eq:lower}
\end{equation}
Therefore, the $\psi$ abundance obeys the limits $3.16 \times
10^{-12} (g\Phi/M_{Pl})^{1/2} \lesssim \Omega_\psi h^2\lesssim
0.14$. This imposes constraints on the $\chi$ bare mass according
to (\ref{omega-psi}):
\begin{equation}
\begin{array}{c}
3.2 \times 10^6 \: g \:
        \ln \left( 4.72 \times 10^{19} \: g^{3/2} \: \left({m_\chi^2 \over M_{Pl}^2}+ 10^{-2}g^2\right)^{1/2} \right)
        \\
\leq  {m_\chi^2 \over M_{Pl}^2 } \leq \\
 3.2 \times 10^6 \: g \:
        \ln \left( 1.49 \times 10^{30} \: g^{3/2} \: \left({m_\chi^2 \over M_{Pl}^2}+ 10^{-2}g^2\right)^{1/4}
\right).
\end{array}
\end{equation}
Such conditions define a parameter space which is convenient to
assess the phenomenological viability of the FKL mechanism as a
generator of $\psi$ non relativistic WIMPzillas. This is possible
because the number density of $\chi$ produced in the
post-inflationary era depends on $g$ and $m_\chi$ and is related to
$\Omega_{\psi}$, so that limits on $\Omega_\psi$ today restrict the
possible values of such parameters. The analyses is summarized in
Figure 1, which can be understood as follows. The upper and lower
limits of the gray area come from the substitution of Eqs.
(\ref{eq:upper}) and (\ref{eq:lower}) in (\ref{omega-psi}). We also
have imposed unitarity constraints on Eq. (\ref{limite}) ($g$ and
$g'\lesssim 1$), so that $3 \times 10^{-3} \lesssim g
\lesssim 1$ (which comprises the upper right triangle in the figure)
and obtained the limits to the left (the minimum value for $g$
given the maximum possible value for $g'$) and to the right (the
maximum value for $g$) in the allowed (dark gray) area in the
figure. For the sake of comparison, we also included the same
analysis for the original FKL result, which corresponds to the
right triangle limited by the dot-dashed line and to the widened
area including the light gray region.

\begin{figure}[th]
\vspace*{5mm}
\centerline{\epsfxsize=3.5in\epsfbox{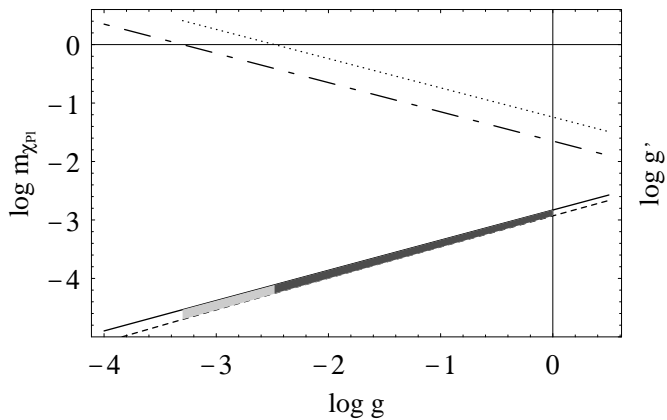}}
\vspace*{9mm}
\caption{The allowed parameter space ($m_\chi,g,g'$) for the
production of WIMPzillas in the FKL mechanism.
For $m_{\chi}$ given in Planck mass units ($m_{\chi_{Pl}}$), the
dark gray area stands for the allowed values of the parameters
between $\Omega_\psi h^2 = 0.14$ (dashed line) and $\Omega_\psi h^2
= 3.16 \times 10^{-12} (g\Phi/M_{Pl})^{1/2}$ (solid line).
Unitarity constraints on $g'$ and $g$ provide limits to the left
and to the right in the allowed area in the figure. The application
of these limits on the original FKL results
\protect\cite{felder}, $g^{\prime 2}g\sim 5 \times 10^{-4}$,
increases the allowed area by adding the light gray region.}
\label{spectrum}
\end{figure}
        We see that the allowed region in the parameter space is rather
constrained. Particularly, given the valid range for $g$, we find
very high masses for $\psi$. From Eq. (\ref{eq:mpsi}) the minimum
$\psi$ mass that can be obtained through this model is $m_\psi \sim
10^{15}$GeV (the upper limit being $m_\psi \sim 10^{18}$GeV. This
happens because the exponential suppression of the number density
of  $\chi$ particles created after the first $\phi$ field half
oscillation, necessary to avoid parametric resonance (see Eq.
(\ref{n2-n1})), is also present in the expression for $\Omega_\psi$ 
(see Eq. (\ref{omega-psi})).
Although the upper cosmological limit is quite strong given the
most recent measurements, one may consider as a weak constraint the
second astrophysical cosmic ray limit, as it is model dependent.
However, even if we consider other classes of models to establish
new lower limits on $\Omega_\psi$ the above results would not
change significantly, because the parameters of the model are not
very sensitive to variations on $\Omega_\psi$. Let us assume, for
the sake of a comparison, an hypothetical value for the present
$\psi$ abundance, say $\Omega_\psi h^2 \sim 10^{-20}$, so that the
allowed region of the Fig. 1 is enlarged. The resulting parameter
space is depicted in Figure 2.
\begin{figure}[th]
\vspace*{5mm}
\centerline{\epsfxsize=3.5in\epsfbox{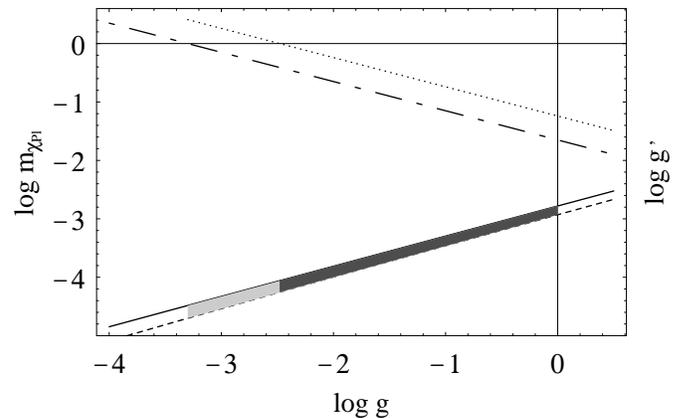}}
\vspace*{9mm}
\caption{The allowed parameter space of the previous figure
is not substantially altered even if it is enlarged by an
exagerated and hypothetical lower limit for the abundance of $\psi$
($\Omega_\psi h^2 =10^{-20}$, solid line). All the other lines have
the same meanings as in Fig. 1}
\end{figure}

For the same reason, wide variations of the reheating temperature
in Eq. (\ref{abundance1}) will not change the picture. For
example, one would have to consider reheating temperatures about 20
orders of magnitude higher than the one assumed here to obtain a
shift of only one order of magnitude on the allowed range of
$m_{\chi_{Pl}}$. Therefore, the resulting allowed area of the
parameter space is relatively independent of particle physics
details of the reheating and of the hypothetical top-down decay of
$\psi$ particles in UHECR's.

\section{Conclusion}

        We studied how to generate supermassive fermions that can explain
the UHECR's in the context of a particle production mechanism
suggested in Ref. \cite{felder}. We obtained the parameter space
for which such a mechanism can take place and concluded that some
fine tuning of parameters seems to be necessary. Additionally, the
lower limit on the $\psi$ mass obtained, $m_\psi \gtrsim 10^{15} $
GeV, is rather robust. A typical signature of this model would be
an unforeseen rise in the flux at the highest energies end of the
cosmic ray spectrum, which could be observed by the next generation
of experiments, like the Pierre Auger observatory. The cosmic ray
spectrum must have a cutoff which is associated to the maximum
energy possible to UHECR's and is independent of the GZK feature.
If such a cutoff happen to be below $10^{15}$ GeV, this simplest
version of the FKL mechanism should be discarded. On the other
hand, it is this mechanism that can provide masses of such
magnitude more naturally than any other top-down versions, so that
if it is at all possible to measure such high energy cosmic rays
and no cutoff in the UHECR's spectrum is observed by the next
generation experiments, this model can become an attractive
candidate. It is also interesting that, despite having perhaps too
many free parameters, this model is rather constrained, and such a
result is relatively insensitive to wide variations of the relevant
cosmological parameters.

        It is important to emphasize that we studied the
production of non relativistic $\psi$ particles only, and the
scenario can be made more complex by considering the production of
$\psi$ particles that are relativistic at the preheating time but
becomes non-relativistic along the Universe evolution. In this
case, the energy transfer from the inflaton field to other fields
may be not very efficient and it would be necessary to consider the
dilution/concentration of $\psi$ particles throughout the several
phases that happened since inflation (coherent oscillations phase,
radiation domination and matter domination) and the allowed
parameter space may be widened.

\acknowledgements The authors are grateful to Hugo C. Reis for
useful discussions. We also benefitted from suggestions and
comments from Raul Abramo, Carlos O. Escobar, Alejandra Kandus,
Gustavo Medina-Tanco and Daniel Muller. This work was supported by
FAPESP and CNPq.

\appendix
\section{The exponential suppression}

        From (\ref{sol.phi-preII}), we can write the time derivative of
$\phi(N)$ with respect to $N$:
\begin{equation}
\phi'(N)= \frac{M_{Pl}}{3} \frac{2\pi\cos{(2\pi N)}}{2 \pi N}-
          \frac{M_{Pl}}{3}\frac{\sin{(2\pi N)}}{2 \pi N^2}.
\end{equation}
By labelling each time that $\phi$ passes through the minimum of
its potential as $N_j$ we have that
\begin{equation}
{\phi'}_j=\frac{M_{Pl}}{3} \frac{\cos{(2\pi N_j)}}{ N_j},
\end{equation}
that is,
\[
\begin{array}{lll}
\mbox{$1^{st}$ passage}: & & \\
\quad N_1  = 1/2 & \rightarrow & |{\phi'}_1|= {2\over 3} M_{Pl}
\nn \\
\mbox{$2^{nd}$ passage}: & & \\
\quad N_2 = 2/2 & \rightarrow & |{\phi'}_2|= {1\over 3} M_{Pl}
\nn \\
\mbox{$3^{rd}$ passage}: & & \\
\quad N_3 = 3/2 \quad & \rightarrow & |{\phi'}_3|= {2\over 9} M_{Pl}
\nn \\
...  &    \nn \\
\mbox{$j^{th}$ passage}: & & \\
\quad N_j = j/2 \quad & \rightarrow & |{\phi'}_j|=\left|\frac{M_{Pl}}{3}
\frac{\cos{(j\pi)}}{j/2}\right|= \frac{M_{Pl}}{3} \times \frac{2}{j}
\end{array}
\]

From the general expression above, it is possible to obtain the
ratio between the particle number density of $\chi$ particles
produced in two consecutive bursts of $\chi$ generation so that we
can evaluate the amount of suppression for each passage $j$.
Writing the expression (\ref{tot.part.number.ip}) for the total
number density of $\chi$ particles produced in each passage in
terms of the new definitions:
\begin{eqnarray}
n_{\chi}^{(j)}(t_j) & = & \frac{g^{3/2}}{8\pi^3}\left(\frac{|{\phi'}_j| m_\phi}{2\pi}\right)^{3/2}
                    \exp{\left(-\frac{\pi m_{\chi}^2}{g|{\phi'}_j m_\phi/2\pi|}\right)}  \nn \\
                & = & {\frac{g^{3/2}}{8\pi^3}\left(\frac{M_{Pl}}{3} \frac{m_\phi}{2\pi j}\right)^{3/2}
                    \exp{\left(-\frac{6\pi^2 j}{M_{Pl}m_\phi} \frac{m_{\chi}^2}{g}\right)}}.
\label{number-j}
\end{eqnarray}
Since the $a(t) \propto t^{2/3}$ along the coherent oscilations
phase, we must compare the number density of particles produced at
$t_j$ until $t_{j+1}$ by taking into account the dilution of $n_j$
at the $(j+1)$th passage:
\begin{eqnarray}
n_{\chi}^{(j)}(t_j)a^3(t_j) & = & n_{\chi}^{(j)}(t_{j+1})a^3(t_{j+1}) \nn \\
\Rightarrow  n_{\chi}^{(j)}(t_{j+1}) & = &
n_{\chi}^{(j)}(t_j)\left(\frac{t_j}{t_{j+1}}\right)^2
\end{eqnarray}
However, from (\ref{N(phi)}), $t_j/t_{j+1}= j/j+1$, so the ratio to
be considered stands:
\begin{equation}
\frac{n_{\chi}^{(j+1)}(t_{j+1})}{n_{\chi}^{(j)}(t_{j+1})}=
\frac{n_{\chi}^{(j+1)}(t_{j+1})}{n_{\chi}^{(j)}(t_j)} \left(\frac{j+1}{j}\right)^2.
\end{equation}
By using the expression (\ref{number-j}) in the above relation, we
finally find:
\begin{equation}
\frac{n_{\chi}^{(j+1)}(t_{j+1})}{n_{\chi}^{(j)}(t_{j+1})}=\left(\frac{j+1}{j}\right)^{1/2}
 \exp{\left(-\frac{6\pi^2 }{M_{Pl}m_\phi} \frac{m_{\chi}^2}{g}\right)}.
 \label{n-sup}
\end{equation}
Since $\chi$ particles decay into two $\psi$'s, the ratio above is
also valid for $\psi$ particles. Thus, we conclude that the
exponential suppression, that has to be addressed in order to obey
cosmological limits on the WIMPzillas abundance (see section IV),
guarantees that we can consider in a good approximation only the
$\chi$ particles produced in the first passage.

\end{multicols}

\end{document}